\documentclass[a4paper,fleqn,usenatbib]{mnras}

\usepackage{newtxtext,newtxmath}

\usepackage[T1]{fontenc}
\usepackage{ae,aecompl}

\usepackage{graphicx}	
\usepackage{amsmath}	
\usepackage{amssymb}	
\usepackage{hyperref}

\title[\textcolor{black}{Structural properties of substructures}]{Statistical properties of substructures around \textcolor{black}{Milky Way-sized} haloes and their implications for the formation of stellar streams}

\author[Yu Morinaga et al.]{
Yu Morinaga,$^{1}$\thanks{E-mail: aefa2627@chiba-u.jp}
Tomoaki Ishiyama,$^{2}$
Takanobu Kirihara$^{2}$
and Kazuki Kinjo$^{1}$
\\
$^{1}$Department of Applied and Cognitive Informatics, Chiba University, 1-33, Yayoi-cho, Inage-ku, Chiba 263-8522, Japan\\
$^{2}$Institute of Management and \textcolor{black}{Information} Technologies, Chiba University, 1-33, Yayoi-cho, Inage-ku, Chiba 263-8522, Japan\\
}

\date{Accepted 2019 May 14. Received 2019 May 10; in original form 2019 January 15}

\pubyear{2018}

\begin{document}
\label{firstpage}
\pagerange{\pageref{firstpage}--\pageref{lastpage}}
\maketitle

\begin{abstract}
Stellar streams originating in disrupted dwarf galaxies and star
clusters are observed around the Milky Way and nearby galaxies. Such
substructures are the important tracers that record how the host
haloes \textcolor{black}{have} accreted progenitor galaxies.  Based on the cosmological
context, we investigate the relationship between structural properties
of substructures such as length and thinness at $z=0$, and orbits of
their progenitors.  We model stellar components of a large sample of
substructures around Milky Way-sized haloes by combining semi-analytic
models with a high-resolution cosmological $N$-body simulation. Using
the Particle Tagging method, we embed stellar components in progenitor
haloes and trace phase-space distributions of the substructures
down to $z=0$.  We find that the length and thinness of substructures
vary smoothly as the redshift when the host haloes accrete their
progenitors.  For substructures observed like streams at $z=0$, a
large part of the progenitors is accreted by their host haloes at
redshift $0.5\lesssim z\lesssim 2.5$.
Substructures with progenitors out of this accretion redshift range
are entirely or less disrupted by $z=0$ and cannot be observed as streams.
We also find that the
distributions of length and thinness of substructures vary smoothly
as pericentre and apocentre of the progenitors.
Substructures observed
like streams tend to \textcolor{black}{have} 
the specific range of
$10\ {\rm kpc} \lesssim r_{\rm peri}\lesssim100\ {\rm kpc}$ and
$50\ {\rm kpc} \lesssim r_{\rm apo}\lesssim300\ {\rm kpc}$.
\end{abstract}




\section{Introduction}
The standard lambda cold dark matter (CDM) scenario predicts that galaxies
are formed via hierarchical mergers of smaller objects over its
lifetime \citep{1978MNRAS.183..341W}.  As a consequence of this merger
process, a large number of substructures such as dwarf galaxies and
their tidally disrupted remnants including stellar streams are
expected to exist around galaxies as fossil records of their accretion
events.

Among many kinds of substructures, stellar streams bring especially
important clues to \textcolor{black}{investigate} galaxy formation.  Infalling dwarf
galaxies and star clusters into their host galaxies are tidally
perturbed, and their stellar components are stripped.
These stripped stellar components are elongated and formed tidal tails approximately
along the progenitor orbits and observed as structures like ``stream'' at present \citep{1996ApJ...465..278J}.

Such streams are the important tracers that record how the host haloes
accreted progenitor galaxies because characteristic structural
properties of streams should be significantly sensitive to their
progenitor orbits \textcolor{black}{\citep[e.g.,][]{2015MNRAS.446.3100H}} and also environments of their host galaxies such as
the shape of underlying gravitational potential and interactions with other substructures \textcolor{black}{\citep[e.g.,][]{1999ApJ...512L.109J, 2013MNRAS.433..878L}}.

So far, a large number of observations have actually discovered
substructures including the Sagittarius stream
\citep[e.g.,][]{1994Natur.370..194I,2001ApJ...551..294I,2003ApJ...599.1082M}, Orphan stream
\textcolor{black}{\citep{2006ApJ...651L..29G,2007ApJ...658..337B,2018arXiv181206066F,2019MNRAS.485.4726K}} around the Milky Way (MW),
the \textcolor{black}{Giant Southern Stream} in the Andromeda galaxy (M31) \citep[e.g.,][]{2001Natur.412...49I} and many
other substructures in nearby galaxies  \citep[e.g.,][]{2008ApJ...689..184M, 2009Natur.461...66M}.  Recent
surveys such as the Sloan Digital Sky Survey (SDSS), the {\it Gaia}
mission \citep{2016A&A...595A...2G}, Dark Energy Survey (DES)\citep{2016MNRAS.460.1270D},
\textcolor{black}{Pan-Andromeda Archaeological Survey (PAndAS) \citep{2009Natur.461...66M, 2013ApJ...776...80M}},
\textcolor{black}{the Panoramic Survey Telescope and Rapid Response System (Pan-STARRS) \citep{2002SPIE.4836..154K} and Subaru Hyper Suprime-Cam \citep{2006SPIE.6269E..0BM,2012SPIE.8446E..0ZM} }
have been discovering new faint dwarf galaxies and stellar streams, and
measuring \textcolor{black}{some properties} of a part of individual
stars \textcolor{black}{\citep[e.g.,][]{2010ApJ...711..671K, 2011ApJ...732...76R, 2012ApJ...752...45T, 2015ApJ...813..109D, 2018Natur.563...85H, 2018ApJ...863...89S, 2018ApJ...862..114S, 2018ApJ...853...29K, 2018ApJ...866...42F, 2018PASJ...70S..18H}}.
These past, ongoing and planning
surveys would enable us to characterise the structural properties for
plenty of substructures such as the length and thinness in
more detail and infer the origin of substructures by
comparing with some theoretical models.

To investigate the relationship between structural properties of
streams, orbits of their progenitors and environments, various studies
have been carried out \textcolor{black}{\citep[e.g.,][]{2001ApJ...551..294I,2001ApJ...557..137J,2008ApJ...689..936J,2003ApJ...599.1082M, 2006ApJ...650L..33P,2010MNRAS.408L..26P, 2007MNRAS.380...15F,2008ApJ...682L..33F,2013MNRAS.434.2779F, 2007MNRAS.379.1464S, 2008MNRAS.389.1391S,2008ApJ...689..184M, 2011MNRAS.417..198V,2012ApJ...748...20C,2014MNRAS.442.3544F, 2014ApJ...783...87M, 2016ApJ...827...82M, 2017MNRAS.470..522S, 2017MNRAS.464.3509K, 2017MNRAS.469.3390K, 2018ApJ...867..101B}}.
\citet{2001ApJ...551..294I} suggested that the gravitational potential
of the MW must be nearly spherical to reproduce the observed
\textcolor{black}{positions} and \textcolor{black}{velocities} of stars in the Sagittarius stream.
A few gaps found along stellar
streams may be caused by a large population of low-mass
subhaloes predicted by the CDM model
\textcolor{black}{\citep{2012ApJ...748...20C,2016ApJ...819....1I,2017MNRAS.470..522S}},
which may be a possible explanation for the so called ``missing satellite'' problem \citep{1999ApJ...522...82K, 1999ApJ...524L..19M}.

For these studies, $N$-body simulations have been extensively used to
reproduce observed properties of substructures and to explore the
dynamical evolution of them.  In many \textcolor{black}{earlier} studies, fixed and
spherical gravitational potential of host galaxies \textcolor{black}{is} assumed, and
orbits of satellite galaxies are free parameters.  However, in the
cosmological context, dynamical evolution of substructures are
significantly affected by complicated physics such as dynamical
friction, multiple interactions with other subhaloes, and evolutionary
history of host galaxies, which could not be treated in fixed
gravitational potential. Orbital parameters of infalling satellites have
some specific distributions \citep{2011MNRAS.412...49W}.
Consequently, tracing the dynamical evolution of substructures using
$N$-body simulations in simple galactic models is insufficient to
estimate their origins precisely.
Studies based on the cosmological context are highly demanded.

\citet{2007ApJ...667..859D} analysed co-evolution
of abundant subhaloes with a host halo formed in their cosmological \textcolor{black}{$N$-body} simulation
and showed the mass evolution of subhaloes depends on orbital
properties such as pericentre.
\citet{2008MNRAS.385.1859W} examined the formation and evolution of
tidal streams originated in cosmological disrupted subhaloes and found
correlations between the structures of streams and some properties of
progenitor haloes such as infalling masses and orbital parameters.
However, the dynamical evolution of subhaloes and dwarf galaxies
differ because tidal stripping preferentially occurs in the outer part
\textcolor{black}{of haloes}.  Besides, they used only one MW-sized host halo, which is not
enough to capture the statistics of structural properties of
substructures.

To shed light on these issues, we explore the dynamical evolution and
structural properties of stellar components of a large sample of
substructures within the cosmological context.  
\textcolor{black}{
In particular, we focus on substructures originating in ``dwarf galaxies'', while \citet{2018arXiv181110084C,2018ApJ...861...69C} have explored the formation and evolution of streams originating in globular clusters and their density gaps.
}
We combine a high-resolution cosmological $N$-body simulation
\citep{2016ApJ...826....9I} with the ``Particle Tagging'' method to embed
stellar components into haloes, which has succeeded to reproduce some
observed features of the stellar halo of the MW
\citep[e.g.,][]{2008MNRAS.391...14D}.
\citet{2010MNRAS.406..744C} have developed an extension method of the Particle Tagging,
which is also used to study stellar streams near solar neighbourhood \citep{2013MNRAS.436.3602G}.
To assign stellar masses to haloes, we use a simple model proposed by
\citet{2009ApJ...696.2179K}.  We aim to investigate the relationship
between structural properties of substructures such as length and thinness
at $z=0$, and the orbits of their progenitors in MW-sized haloes.

This paper is organised as follows.  In Section \ref{Cosmological nbody simulation}, we describe the details of our cosmological
$N$-body simulation.  In Section \ref{Methods}, we explain the
analytic models and \textcolor{black}{the definition of} structural and orbital properties of
substructures.  In Section \ref{Results}, we show the results of our
statistical analysis of substructures.
Finally, we discuss and summarise our results in Section \ref{Summary}.

\section{Cosmological $N$-body Simulation}
\label{Cosmological nbody simulation}
We used a high-resolution cosmological $N$-body simulation conducted by \citet{2016ApJ...826....9I}.
This simulation consists of $2048^3$ dark matter particles in an $11.8\ {\rm Mpc}$ comoving cubic box.
The particle mass is $7.54\times10^3M_{\sun}$,
and the gravitational softening length  is $\textcolor{black}{\varepsilon=}176.5\ {\rm pc}$.
The cosmological parameters are consistent with the observation of the cosmic microwave background obtained by the Planck satellite \citep{2014A&A...571A..16P,2018arXiv180706209P}, namely, $\Omega_0=0.31, \Omega_{\rm b}=0.048, \lambda_0=0.69, h=0.68, n=0.96$, and $\sigma_8=0.83$.
\textcolor{black}{The snapshots were stored at the redshifts so that the logarithmic interval $\Delta {\rm log}(1+z)=0.01$.}
We identified dark matter haloes and subhaloes using ROCKSTAR (Robust Overdensity Calculation using K-Space Topologically Adaptive Refinement) halo/subhalo finder \citep{2013ApJ...762..109B}.
Then we constructed their merger trees using CONSISTENT TREE code \citep{2013ApJ...763...18B}, 
Further details of this simulation are given in \citet{2016ApJ...826....9I}.

In this study, we analysed nine MW-sized haloes (GX1-GX9).  Their
virial mass is between $0.55-2.84\times10^{12}M_{\sun}$, 
\textcolor{black}{where the definition of virial mass is given by \citet{1998ApJ...495...80B}}.
Their properties such as the virial mass $M_{\rm vir}$ and the virial radius
$R_{\rm vir}$ are summarised in Table~\ref{tab:tab1}.  From their
merger trees, we selected progenitor haloes with masses more massive
than $M_{\rm vir}>10^7M_{\sun}$ at redshift $z_{\rm acc}$ when they
first pass through the virial radius of the most massive progenitors
of MW-sized haloes (so-called ``main-branch'').  Then, we traced the
evolution of these haloes after the redshift $z_{\rm acc}$.  The
number of subhaloes in each host halo is listed in
Table~\ref{tab:tab1}.

After the redshift $z_{\rm acc}$, most progenitor haloes orbit host haloes as subhaloes.
Some of them are disrupted by the gravitational interaction with the host haloes and \textcolor{black}{cannot} be observed at $z=0$ as subhaloes in the merger tree.
In the following section, we tag a part of member particles of subhaloes at $z_{\rm acc}$ using the ``Particle Tagging'' method and regard these particles as the stellar component.
By tracing these stellar particles down to $z=0$, we can analyse the phase space distributions of both \textcolor{black}{surviving} and disrupted ``galaxies'' at $z=0$.
\textcolor{black}{Hereafter, including streams, we refer to these objects as ``substructures''.
In Section~\ref{definition of structural properties}, 
we categorise them by length and thinness of substructures at $z=0$ into three types, 
self-bounded subhalo, stream and disrupted substructure. 
Because there is no consensual and rigid definition of them from the viewpoint of observations,
we refer to substrcutures using these three terms as a matter of convenience.
}
\begin{table*}
\caption{
Properties of nine Milky-Way sized haloes.
$M_{\rm vir}$ and $R_{\rm vir}$ are
the virial mass and radius.
$N_{\rm sub}$ is the number of subhaloes with $M_{\rm vir}(z_{\rm acc}) >10^7M_{\sun}$,
where $z_{\rm acc}$ is the redshift when progenitors of substructures first pass through the virial radius of
the most massive progenitors of their MW-sized host haloes.
$N_{{\rm sub},M_*=10^{4-5}},\ N_{{\rm sub},M_*=10^{5-6}}$ and $N_{{\rm sub}, M_*>10^{6}}$ are the number of substructures
whose stellar mass ranges are $M_*=10^{4-5},10^{5-6}$ and $10^{6-}M_{\sun}$, respectively.
The numbers in the brackets are those of the streams with each stellar mass ranges.
}

\centering
\begin{tabular}{ccccccc}
\hline
Name & $M_{\rm vir}$ & $R_{\rm vir}$ & $N_{\rm sub}$ & $N_{{\rm sub},M_*=10^{4-5}}$ & $N_{{\rm sub},M_*=10^{5-6}}$ & $N_{{\rm sub},M_*>10^{6}}$

 \\
 \ \ \ & {\scriptsize $[\times10^{12}M_{\sun}]$} & {\scriptsize [${\rm kpc}$]} & & & &\\
 \hline
 \hline
 GX1 & $2.84$ & $372$ & 6560 & 158 (16) & 52 (8) & 18 (2)\\
 GX2 & $2.36$ & $350$ & 4956 & 104 (23) & 48 (8) & 20 (2)\\
 GX3 & $2.27$ & $345$ & 3810 &\ 86 (11) & 20 (2) & 14 (2)\\
 GX4 & $1.09$ & $270$ & 2421 &\ 69\ \ \ (6)\  & 19 (0) &  9 (1)\\
 GX5 & $0.74$ & $238$ & 1853 &\ 32\ \ \ (7)\  & 20 (5) &  7 (0)\\
 GX6 & $0.45$ & $201$ & 959 &\ 17\ \ \ (0)\  & 11 (0) &  5 (0)\\
 GX7 & $0.59$ & $221$ & 1135 &\ 35\ \ \ (2)\  & 8 (1) &  6 (0)\\
 GX8 & $0.60$ & $222$ & 1421 &\ 26\ \ \ (2)\  & 11 (3) &  9 (0)\\
 GX9 & $0.55$ & $216$ & 1016 &\ 26\ \ \ (1)\  & 11 (0) &  3 (0)\\
 \hline
\end{tabular}
\label{tab:tab1}
\end{table*}

\section{Methods}
\label{Methods}

\subsection{Particle Tagging}
\label{Particle Tagging}

Adopting the ``Particle Tagging'' method
\citep[e.g.,][]{2001ApJ...548...33B,2005MNRAS.364..367D,
  2005ApJ...635..931B, 2008MNRAS.391...14D, 2010MNRAS.406..744C} to
dark matter only simulations, we can embed stellar components in
progenitor haloes and trace their phase space distributions down
to $z=0$.  Most stellar components are assumed to be formed in the
centre of progenitor haloes until the redshift $z_{\rm acc}$ when they
first pass through the virial radius of their host haloes.
We tag \textcolor{black}{a} fixed fraction of $f_{\rm MB}$ of the most bound particles of those
progenitors at $z_{\rm acc}$,\footnote{\textcolor{black}{If the progenitor halo is a ``phantom'' halo \citep{2013ApJ...762..109B} in the merger tree at $z_{\rm acc}$, we trace the progenitor back to redshift when it is not phantom and \textcolor{black}{perform} the tagging.}} and treat these particles as ``stellar particles''.  Tracing these particles down to $z=0$, we can investigate
the phase space distributions of substructures originating in
progenitor haloes.  We set $f_{\rm MB}=0.10$ by default.

The similar approach was adopted by
\citet{2008MNRAS.391...14D}.
They also used $f_{\rm MB}=0.10$ and showed that physical properties such as the metallicity and \textcolor{black}{the} age
of stars in accreted stellar haloes of their model galaxies \textcolor{black}{were good agreement with the observed \textcolor{black}{data}.}
Besides, they found that observed structural properties of \textcolor{black}{the stellar component around the MW} were \textcolor{black}{well} reproduced,
reinforcing the effectiveness of the Particle Tagging method.
\citet{2010MNRAS.406..744C} have also developed an extension of this method\textcolor{black}{, which is confirmed to be able to provide} an excellent approximation to hydrodynamical simulations \citep{2017MNRAS.469.1691C}.

In our study, combining the Particle Tagging method with the higher resolution
cosmological $N$-body simulation described in \textcolor{black}{Section} \ref{Cosmological nbody simulation},
we can resolve substructures such as streams originating in smaller haloes
and investigate structural properties of them.
Although we used the fraction $f_{\rm MB}=0.10$ as a fiducial value, we
also compared \textcolor{black}{the} results with $f_{\rm MB}=0.05$ and 0.20, and
confirmed that the differences in $f_{\rm MB}$ do not strongly affect
statistical results of the relationship between structural properties of
substructures and orbits of their progenitors. The detail is given
in Appendix \ref{appendix:1}.

\subsection{A model to assign stellar masses to haloes}
In cosmological $N$-body simulations, MW-sized haloes contain a number
of subhaloes as \textcolor{black}{listed} in Table \ref{tab:tab1}.  However, the number of
known dwarf galaxies in the MW and M31 is two or three dozens
\citep{2012AJ....144....4M}.  This disagreement is known as the
so-called ``missing satellite'' problem \citep{1999ApJ...522...82K,
  1999ApJ...524L..19M}.  To investigate the statistical properties of
visible substructures, it is necessary to assign stellar masses to progenitor
haloes in a physically motivated manner.

Photoionising by \textcolor{black}{the cosmic UV background radiation} sufficiently suppresses the star
formation in low-mass haloes with \textcolor{black}{the} viral temperatures $T_{\rm vir}<10^4\ {\rm K}$. Such haloes unable to cool the gas and form
stars even if sufficient amount of gas exists \citep{1997ApJ...476..458H}.
To assign stellar masses to subhaloes,
we used a model based on this picture proposed by
\citet{2009ApJ...696.2179K} that reproduces the distribution of dwarf
galaxies observed by the Sloan Digital Sky Survey.

In this model, when the circular velocity of a progenitor halo $V_{\rm circ}$ at \textcolor{black}{the} reionization epoch $z_{\rm rei}\sim11$ \textcolor{black}{\citep[e.g.,][]{2009ApJS..180..306D}} is above a critical
threshold $V_{\rm crit,r}\sim10\ {\rm km}\ {\rm s^{-1}}$
(corresponding to $T_{\rm vir}\sim10^4\ {\rm K}$), the stellar mass of
the progenitor halo is given by Equation
(\ref{eq:suppression_model_1})
\begin{eqnarray}
\label{eq:stellar_mass}
\scalebox{0.97}{$M_{*}=\frac{f_{*}(M_{\rm sat}-M_{\rm rei})}{(1+0.26(V_{\rm crit}/V_{\rm circ}(z_{\rm acc}))^3)^3}+f_{*}M_{\rm rei}\ \ (V_{\rm circ}(z_{\rm rei})>V_{\rm crit,r})$},
\label{eq:suppression_model_1}
\end{eqnarray}
where $M_{\rm sat}$ and $M_{\rm rei}$ are the virial masses of \textcolor{black}{the} progenitor halo at $z_{\rm acc}$ and $z_{\rm rei}$, respectively, and the stellar mass fraction is $\textcolor{black}{f_*=}10^{-3}\times\Omega_{\rm 0}/\Omega_{\rm b}$.
In this case, it is assumed that such halo is massive enough to form stars in the pre-reionization era.
On the other hand, for a progenitor halo with $V_{\rm circ}(z_{\rm rei})<V_{\rm crit,r}$, the stellar mass is assigned by Equation (\ref{eq:suppression_model_2}),
assuming very low star formation efficiency in the pre-reionization era\textcolor{black}{,}
\begin{eqnarray}
\label{eq:stellar_mass}
 M_*=\frac{f_{*} M_{\rm sat}}{(1+0.26(V_{\rm crit}/V_{\rm circ}(z_{\rm acc}))^3)^3} & (V_{\rm circ}(z_{\rm rei})<V_{\rm crit,r})\textcolor{black}{.}
 \label{eq:suppression_model_2}
\end{eqnarray}

In these equations, the suppression of star formation occurs for
haloes with the circular velocity below a critical value $V_{\rm crit}$
after the reionization, based on cosmological hydrodynamical simulations
\citep[e.g.,][]{2000ApJ...542..535G, 2006MNRAS.371..401H, 2008MNRAS.390..920O}
, and thus resulting stellar masses are significantly affected by the
choice of $V_{\rm crit}$.
\citet{2000ApJ...542..535G} proposed the critical circular velocity
$V_{\rm crit}\sim40\ {\rm km\ s^{-1}}$,
but lower critical values $V_{\rm crit}\sim20-25\ {\rm km\ s^{-1}}$ were \textcolor{black}{suggested} by
\citet{2006MNRAS.371..401H} and \citet{2008MNRAS.390..920O}.
We \textcolor{black}{vary} $V_{\rm crit}$ and select an appropriate value
so that the observed stellar mass function of dwarf galaxies \textcolor{black}{is} reproduced well.

\begin{figure*}
	\includegraphics[width=170mm]{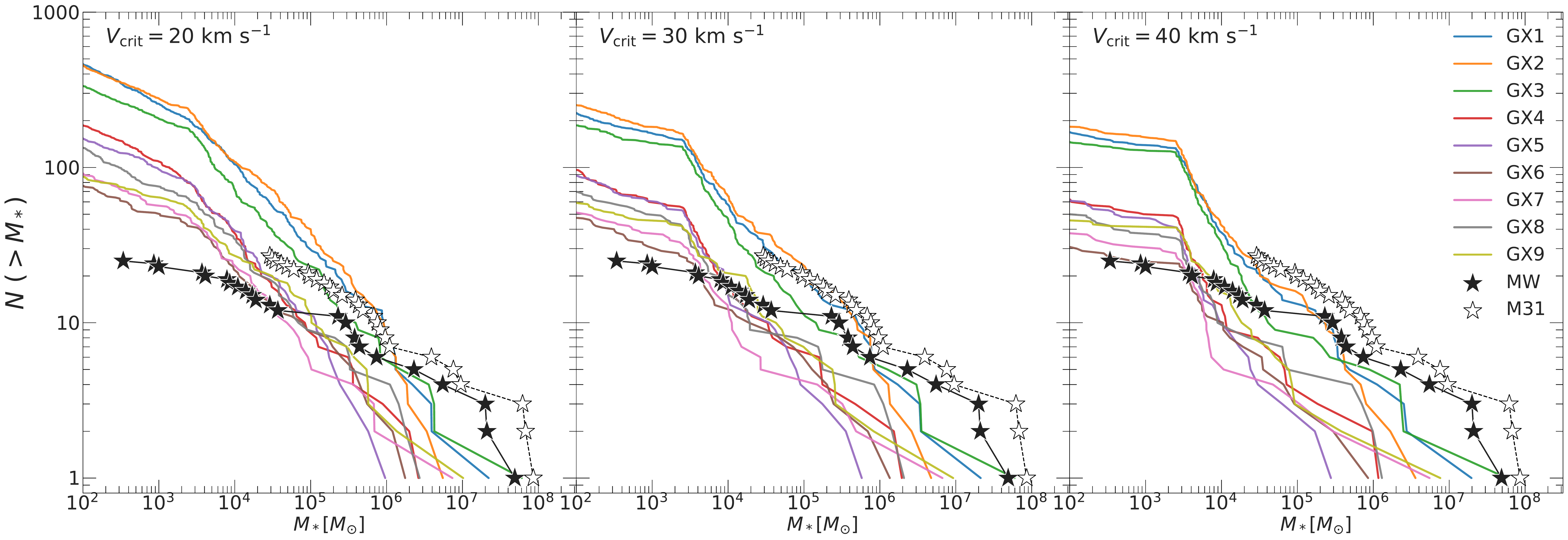}
    \caption{
Cumulative number of subhaloes at $z = 0$ in nine MW-sized host haloes
(GX1-GX9) as a function of their stellar mass, for models with $V_{\rm crit}
= 20,30$ and $40\ {\rm km\ s^{-1}}$. Filled and open \textcolor{black}{symbols} are respectively those of the observed dwarf galaxies in the MW and M31, \citep{2012AJ....144....4M}, \textcolor{black}{and the stellar masses are derived from absolute visual magnitude assuming a stellar-to-light ratio of one}.
The data of Large/Small Magellanic Clouds, \textcolor{black}{M33 and M32} are excluded.
}
    \label{fig:hist1d_2}
\end{figure*}

Figure~\ref{fig:hist1d_2} shows the cumulative number of subhaloes at
$z=0$ (not substructures) in nine MW-sized host haloes (GX1-GX9) as a
function of stellar mass, for models of $V_{\rm crit}=20,30$ and
$40\ {\rm km\ s^{-1}}$.
Filled and open symbols are the distributions of the observed dwarf
galaxies in the MW and M31 \citep{2012AJ....144....4M}, \textcolor{black}{respectively}.
\textcolor{black}{\citet{2009ApJ...696.2179K} adopted this model and} successfully reproduced the distribution of the observed stellar mass function of dwarf galaxies excluding Large/Small Magellanic Clouds.
\textcolor{black}{In the same manner, we also plot the stellar mass function of the MW and M31 excluding the Magellanic Clouds, M33 and M32.}

Subhaloes with \textcolor{black}{lower} stellar masses ($M_*<10^4M_{\sun}$) are more
abundant than observed dwarf galaxies of both the MW and M31
regardless of $V_{\rm crit}$.  One of the \textcolor{black}{reasons} is that it is hard to
observe such \textcolor{black}{ultra} faint dwarf galaxies \textcolor{black}{due to the detection limit.
However a part of such faint dwarf galaxies} would be discovered by ongoing deep imaging surveys by Subaru Hyper Suprime-Cam \citep[e.g.][]{2016ApJ...832...21H,2018PASJ...70S..18H}.
Another reason is that the suppression of star formation in low-mass haloes
might be insufficient in this model. 
\textcolor{black}{
Because such dwarf galaxies are too faint, we only  
analyse substructures with $M_*>10^4M_{\sun}$ in this study.
}
Thus, this disagreement of the number of low
stellar mass dwarf galaxies
\textcolor{black}{does not change our conclusion.}

\textcolor{black}{Although} subhaloes with the stellar mass $M_*>10^8M_{\sun}$ are not found in any host haloes in our simulation,
\textcolor{black}{such satellites} are observed in the MW and M31 such as Large/Small Magellanic Clouds, \textcolor{black}{M33} and M32.
Recent cosmological simulations have also shown that MW-sized haloes to host such massive satellites
are rare \citep{2011MNRAS.414.1560B,2010ApJ...710..408B}.
Therefore, this disagreement does not affect the statistical properties of substructures.
When we exclude these dwarf galaxies, the agreement between our model and the observation \textcolor{black}{becomes} better for $V_{\rm crit}=20$ and $30\ {\rm km\ s^{-1}}$.
Hereafter, we use $V_{\rm crit}=30{\rm\ km\ s^{-1}}$, \textcolor{black}{and} the number of substructures (not subhaloes at $z=0$) with the critical threshold $V_{\rm crit}=30{\rm\ km\ s^{-1}}$ in each host halo is listed in Table~\ref{tab:tab1}.

\subsection{Orbital parameters}

To analyse orbital histories of subhaloes and how they
relate to the properties of substructures,
we quantify the pericentre and apocentre of subhaloes.
\citet{2011MNRAS.412...49W} explored orbital properties of
infalling satellites and reported \textcolor{black}{their dependence} on
the host and satellite masses and redshift.  They calculated the
orbital circularity and pericentre by the two-body approximation (host and satellite haloes).  
This procedure would be valid for infalling
satellites.  However, after $z_{\rm acc}$, actual orbital properties
such as pericentre and apocentre evolve via dynamical friction, tidal disruption, and multiple interactions between subhaloes.  Therefore,
the two-body approximation is not accurate enough to describe \textcolor{black}{the} orbital properties.

To calculate the pericentre and apocentre of subhaloes
more accurately, we
trace their orbits from $z_{\rm acc}$ to $z=0$ or \textcolor{black}{the}
redshift when they are entirely merged with their host haloes.
Then we define the pericentre (apocentre) as the smallest value
in local minima (maxima) of the radial distance of subhaloes.
When subhaloes do not experience any pericentre or apocentre passages, we define the pericentre as the smallest radial distance and the apocentre as the largest radial distance after $z_{\rm acc}$ so that we can quantify \textcolor{black}{the orbital properties}.

\subsection{Quantifying structural properties of substructures}
\label{definition of structural properties}
We characterise structural properties of substructures at $z=0$ and explore the relationship between them and the orbits of progenitors.
We quantify them with two characteristic parameters, length $L_{\rm sub}$, and thinness $T_{\rm sub}$.

\subsubsection{Definition of the length of substructures $L_{\rm sub}$}
\label{definition of length}

\textcolor{black}{
  We define the length of substructures as follows.
  At first, we perform the Principal Component Analysis (PCA) for the stellar particles of substructures in the three-dimensional coordinate.
  The spatial coordinate is transformed to the new coordinate (PC1-3).
  Then, we define the length of a substructure $L_{\rm sub}$ as the sum of physical distances of two stellar particles that lie furthest along the PC1 axis to the centre of the substructure.
  To exclude outliers, we count the number of stellar particles in the substructure along the PC1 axis at intervals of $15\varepsilon\ (=15\times176.5\ {\rm pc})$ and remove particles in the region whose number of particles is below five in calculating the length.
  In this case, the threshold of the outlier removal step is represented by $\rho_{N_{\rm p}}=5$.
  Although we use $\rho_{N_{\rm p}}=5$ as a fiducial value throughout this paper, we also compare the results of the structural properties of substructures with $\rho_{N_{\rm p}}=0\ {\rm and}\ 25$ in Appendix \ref{appendix:2}, 
which indicates that statistical results are insensitive to the choice of $\rho_{N_{\rm p}}$.
  }

\subsubsection{Definition of the thinness of substructures $T_{\rm sub}$}
\label{definition of thinness}
To quantify the thinness of substructures, we apply a similar method proposed by \citet{2017MNRAS.470..522S}.
\textcolor{black}{We perform PCA for the stellar particles of substructures in the three-dimensional coordinate, and each stellar particle (denoted by $i$) is represented by principal components scores $t_{k(i)} (k = 1, 2, 3)$.}
The PC1 (PC3) scores have the largest (smallest) variance in the three.

Using these variances, we define the thinness $T_{\rm sub}$ as
\begin{eqnarray}
\label{eq:thinness}
T_{\rm sub}=\sqrt{\frac{\sum_{i=1}^{N_{\rm p}}\left({\bf t_{\rm 1({\it i})}}-\overline{\bf{t_{\rm 1}}}\right)^2}{\sum_{i=1}^{N_{\rm p}}\left({\bf t_{\rm3({\it i})}}-\overline{\bf{t_{\rm 3}}}\right)^2}} \quad ,
\end{eqnarray}
where $N_{\rm p}$ is the number of stellar particles in the
substructure.  In other words, $T_{\rm sub}$ represents the ratio of
the standard deviation of PC1 to PC3.  For example, a substructure
with high-$T_{\rm sub}$ is elongated along PC1 and would be observed
like a stream at $z=0$.  On the other hand, a substructure with
low-$T_{\rm sub}\ (\sim 1)$ distributes three-dimensionally, which
means it is entirely disrupted or is not much tidally affected.

\subsubsection{Definition of the stream}
\label{definition of the stream}
We consider that substructures with large values of the length and thinness can be observed as stellar streams.
In this paper, we refer to substructures with $L_{\rm sub}/R_{\rm vir}>5$ and $T_{\rm sub}>6$ at $z=0$ as streams,
where $R_{\rm vir}$ is the virial radius of the progenitor halo at $z_{\rm acc}$.
In the case of massive substructures, $L_{\rm sub}$ is naturally high even if they are not tidally disrupted.
To pick up tidally elongated substructures, we define the stream by $L_{\rm sub}/R_{\rm vir}$,
which represents the relative disruption of substructures.
Entirely disrupted substructures or slightly disrupted substructures are not categorised as streams in this definition
because such substructures have smaller values of $T_{\rm sub}$.

\section{Results}
\label{Results}

\subsection{Distribution of length $L_{\rm sub}$ and thinness $T_{\rm sub}$}

\begin{figure}
\includegraphics[width=80mm]{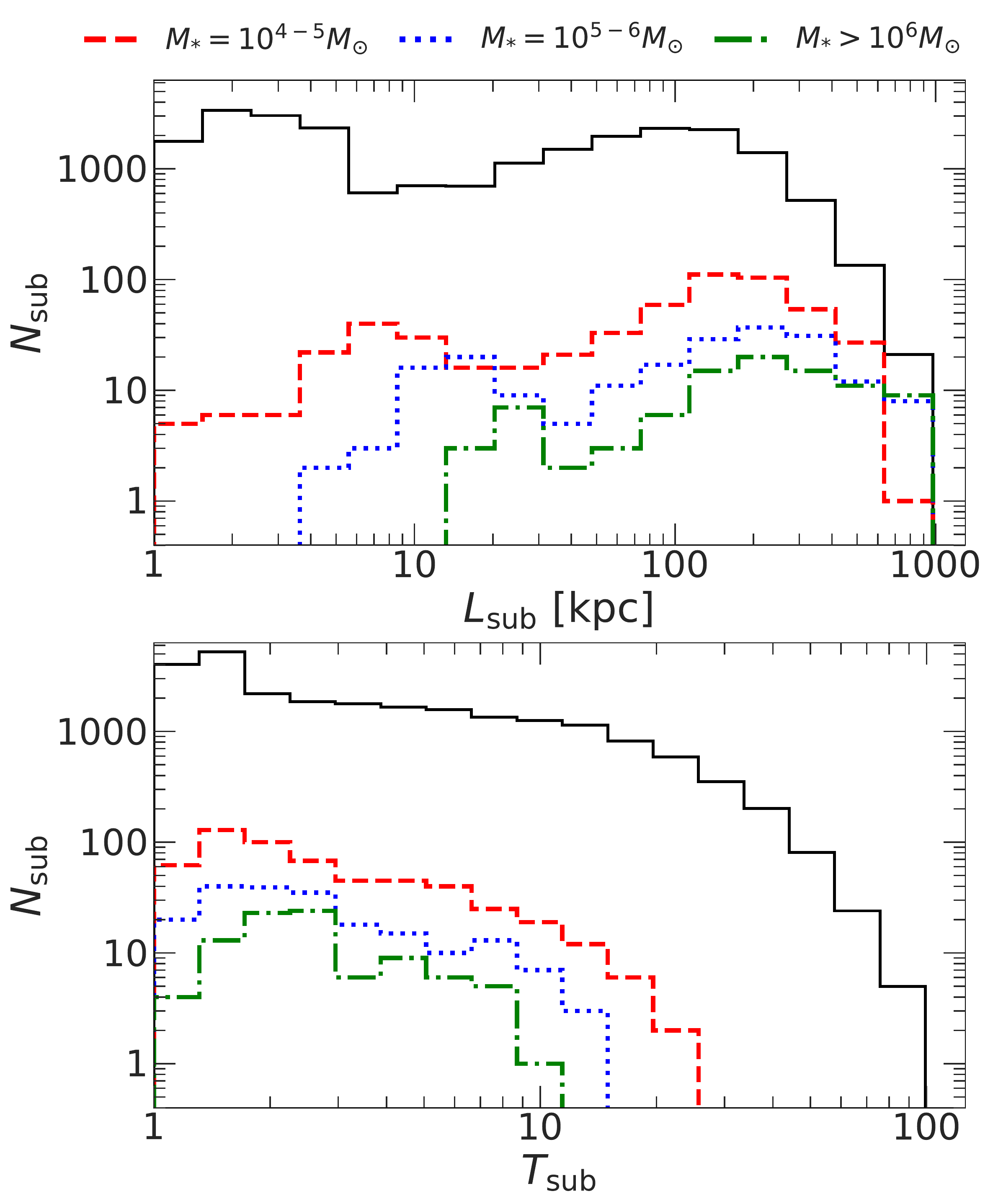}
\caption{
Distribution of length $L_{\rm sub}$ (top panel) and thinness
$T_{\rm sub}$ (bottom panel) of \textcolor{black}{all} substructures \textcolor{black}{in nine MW-sized haloes (GX1-GX9)} at $z=0$.
\textcolor{black}{\textcolor{black}{Red} dashed, blue doted and green dash-dotted curves}
show the results of substructures with the stellar mass ranges of
$M_*=10^{4-5},10^{5-6}$ and $10^{6-}M_{\sun}$, respectively.
Solid curves show the distribution of substructures with
$M_{\rm  vir}(z_{\rm acc})>10^7\ M_{\sun}$.
}
\label{fig:hist1d_structure}
\end{figure}

Figure~\ref{fig:hist1d_structure} shows the distributions of length and
thinness of \textcolor{black}{all} substructures with stellar mass ranges of
$M_*=10^{4-5},10^{5-6}$ and $10^{6-}M_{\sun}$ \textcolor{black}{in nine MW-sized haloes (GX1-GX9)}.
The distribution of the length shows a bimodality regardless of the stellar mass range.  The first
peak at short length around $10$ kpc originates from less disrupted
substructures.  The second peak at the long length from $100$ to
$200\ {\rm kpc}$ originates from entirely disrupted substructures.
The first peak slightly shifts towards the long length with
increasing the stellar mass because the size of progenitor haloes
becomes larger.  On the other hand, the second peak is almost
unchanged.

The distribution of thinness shows that the number of substructures
decreases as the thinness increases\textcolor{black}{,} and \textcolor{black}{peaks} at around $1$.  This
trend is also seen in substructures with any stellar mass ranges,
indicating that the highly elongated substructures are quite rare and
a large part of substructures is not observed as streams at $z=0$.

\subsection{Relation between length \textcolor{black}{$L_{\rm sub}$}, thinness \textcolor{black}{$T_{\rm sub}$} and accretion redshift $z_{\rm acc}$}
\begin{figure}
\includegraphics[width=80mm]{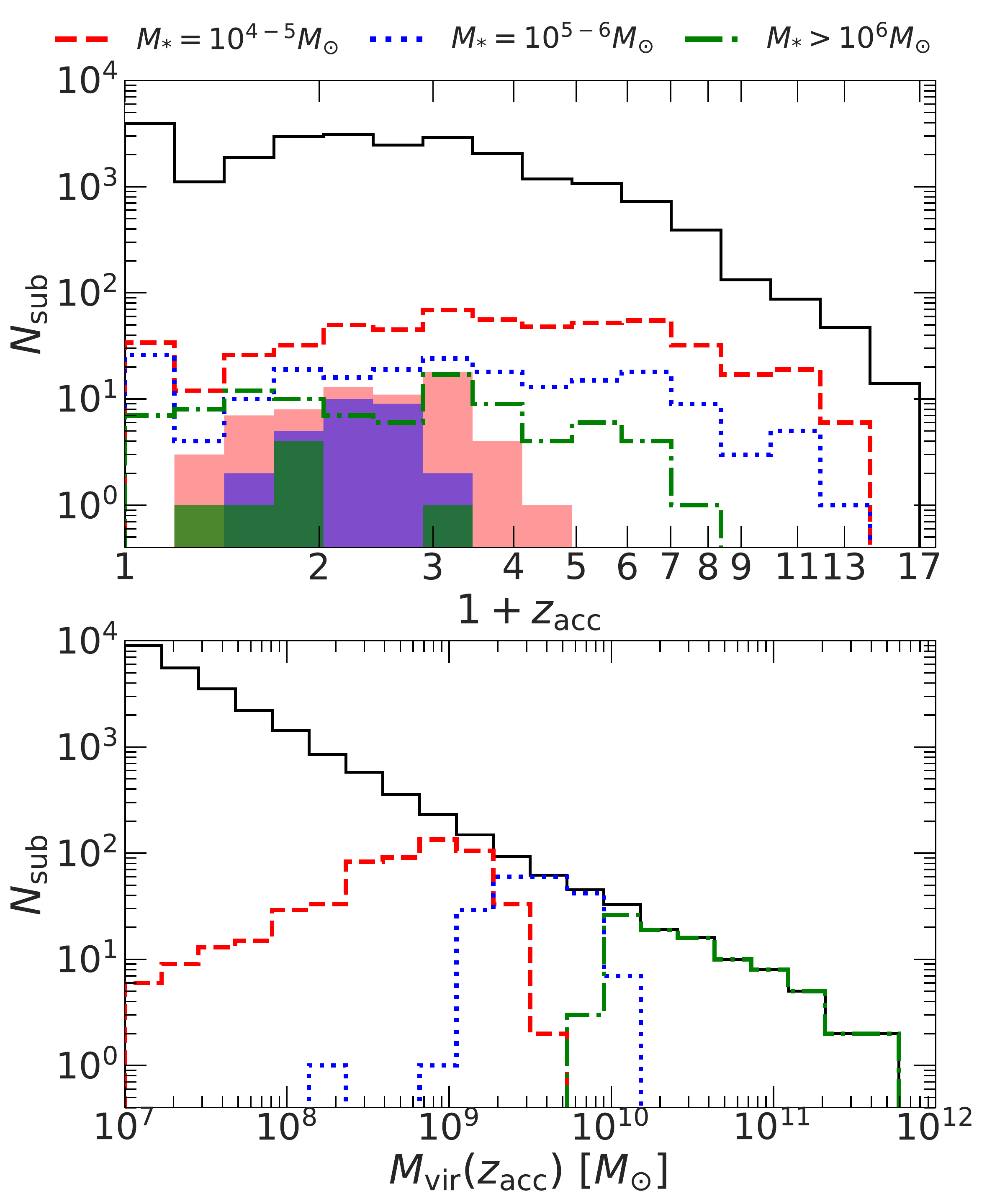}
\caption{
Top: Histograms of $z_{\rm acc}$,
when progenitors of substructures first pass through the virial radius of
the most massive progenitors of their MW-sized host haloes.
Bottom: Histograms of the virial mass of progenitors of substructures
$M_{\rm vir}(z_{\rm acc})$ at $z_{\rm acc}$.
Dashed curves show the distribution of substructures
with the stellar mass ranges of
$M_*=10^{4-5},10^{5-6}$ and $10^{6-}M_{\sun}$, respectively.
Solid curves show the distribution of substructures
with $M_{\rm  vir}(z_{\rm acc})>10^7\ M_{\sun}$.
In the upper panel, the filled histograms
show the distributions of $z_{\rm acc}$ of streams,
which are defined in \S \ref{definition of the stream}.
}
\label{fig:hist1d_Zacc_Macc}
\end{figure}

To explore the accretion histories of substructures around MW-sized
haloes at $z=0$, we plot the distributions of $z_{\rm acc}$ and masses
of their progenitor haloes $M_{\rm vir}(z_{\rm acc})$ at $z_{\rm acc}$
in Figure~\ref{fig:hist1d_Zacc_Macc}.  From the overall distribution
of $z_{\rm acc}$, the number of progenitors with mass
$M_{\rm vir}(z_{\rm acc})>10^7M_{\sun}$ tends to be decreasing with
increasing redshift from $z=4$, just because haloes more massive than
this value are not enough formed (the half mass formation epoch of
such haloes is $z<3$ \citep{2015PASJ...67...61I}). These trends
propagate substructures with the stellar mass $M_*>10^4M_{\sun}$.
The distributions of streams are clearly different from the overall
distribution of substructures.  A large part (approximately $90\%$) of
the streams is accreted by their host haloes within $0.5\lesssim
z_{\rm acc}\lesssim2.5$.  On the other hand, only $20\%$ of all
substructures is accreted within this redshift range.

As shown in the bottom panel of Figure~\ref{fig:hist1d_Zacc_Macc}, the
mass distribution of substructures with the progenitor mass $M_{\rm vir}(z_{\rm
  acc})>10^7M_{\sun}$ approximately follows a power law.  However,
those within a certain range of stellar masses differ depending on the
range.  The substructures with the stellar mass ranges of
$M_*=10^{4-5},10^{5-6}$ and $10^{6-}M_{\sun}$ approximately
correspond to the progenitor haloes with the mass ranges of $M_{\rm vir}(z_{\rm acc})=10^{7-9},10^{9-10}$ and $10^{10-}M_{\sun}$,
respectively.  Almost all the progenitor haloes with mass $M_{\rm vir}(z_{\rm acc})\gtrsim10^9M_{\sun}$ have relatively massive stellar components ($M_*>10^{\textcolor{black}{5}}M_{\sun}$).
On the other hand, in low-mass progenitor haloes, the star formation is strongly suppressed in our model due to \textcolor{black}{the} photoionised heating by the cosmic UV background \textcolor{black}{radiation}.
Therefore, a large part of low-mass progenitor haloes with the mass below $10^8M_{\sun}$ at $z_{\rm acc}$ \textcolor{black}{is} excluded in our analysis.

\begin{figure*}
\includegraphics[width=160mm]{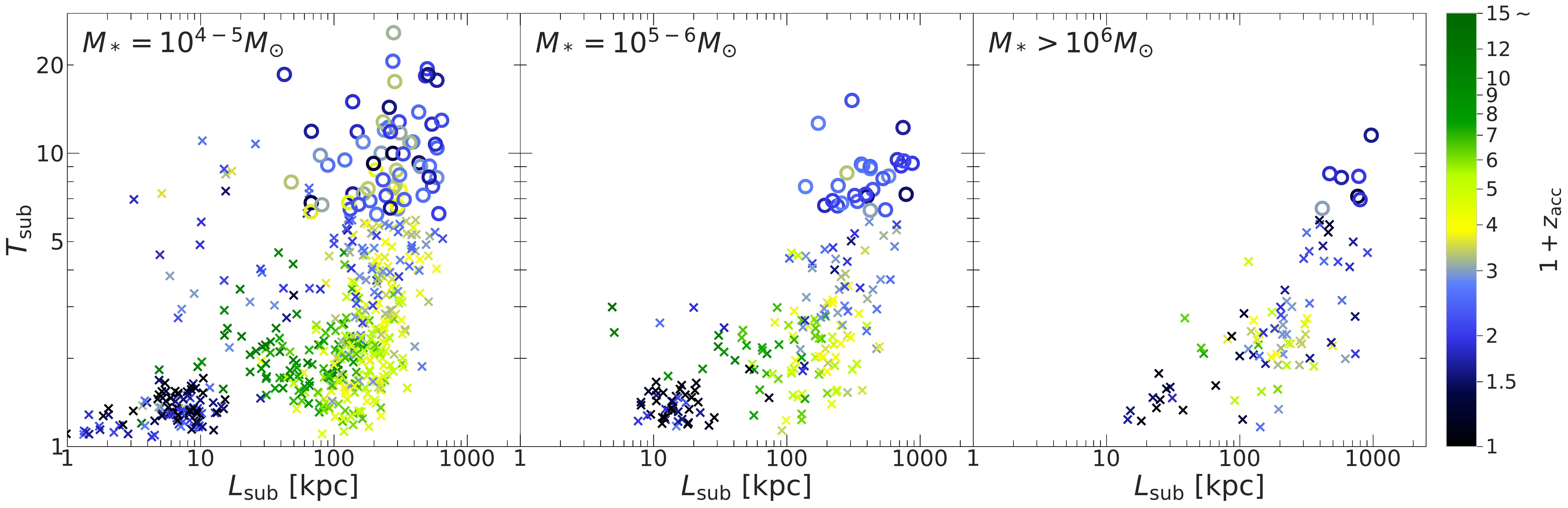}
\caption{
Distributions of length $L_{\rm sub}$ and thinness $T_{\rm sub}$ for
substructures with stellar mass ranges of
$M_*=10^{4-5},10^{5-6}$ \textcolor{black}{and} $10^{6-}M_{\sun}$.
The symbol colour gives the \textcolor{black}{value} of $z_{\rm acc}$. Circles denote the streams and crosses denote
the others.
}
\label{fig:hist3d_zacc}
\end{figure*}

Figure~\ref{fig:hist3d_zacc} shows the distributions of length and
thinness at $z=0$ for substructures with the stellar
mass ranges of $M_*=10^{4-5},10^{5-6}$ and $10^{6-}M_{\sun}$.
The length tends to get longer,
and the thinness tends to get slightly smaller for \textcolor{black}{substructures} with more stellar \textcolor{black}{masses} as
shown in Figure~\ref{fig:hist1d_structure}.
The length and thinness vary smoothly as the accretion redshift $z_{\rm acc}$.
A large part of substructures with $z_{\rm acc}\lesssim0.5$ exists in
the region with the short length ($L_{\rm sub}\sim10\ {\rm kpc}$) and \textcolor{black}{the}
small thinness ($T_{\rm sub}\sim1$), indicating that these
substructures are less affected by tidal forces from their host haloes
because of recent accretion (small $z_{\rm acc}$).
On the other hand, a large part of substructures with $z_{\rm acc} \gtrsim 0.5$ is strongly perturbed by the tidal force and has the long length.
The typical length of substructures at $z_{\rm acc}$ is about 1-10 kpc, increased by a factor of approximately 1-50 until $z=0$.
These substructures show a clear correlation between $T_{\rm sub}$ and $z_{\rm acc}$.
The thinness is decreasing with increasing $z_{\rm acc}$. In particular, substructures with the long length
($ L_{\rm sub} \gtrsim 100\ {\rm kpc} $) and \textcolor{black}{the} large thinness ($T_{\rm sub} \gtrsim 6$), which are defined as streams in this work, give a
specific redshift range of $0.5\lesssim z_{\rm acc}\lesssim2.5$
as shown in Figure~\ref{fig:hist1d_Zacc_Macc}.

\begin{figure*}
\includegraphics[width=170mm]{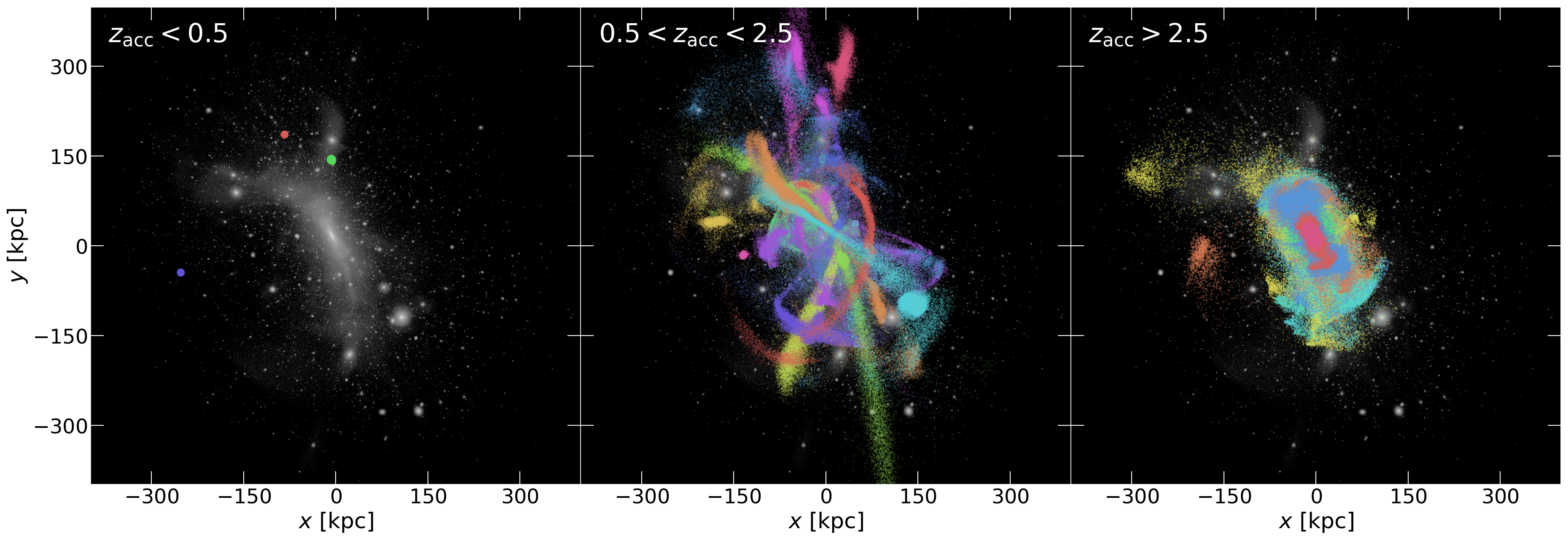}
\caption{
Distributions of stellar particles of substructures $M_{*}=1-6\times10^5M_{\sun}$
around a MW-sized halo (GX1) at $z=0$.
Each panel plots substructures  originating in progenitors with different
accretion redshift ranges of $z_{\rm acc}<0.5, 0.5< z_{\rm acc}<2.5$ \textcolor{black}{and} $z_{\rm acc}> 2.5$.
Each progenitor is visualised by different colours.
The background image of each panel shows
the projected dark matter density distributions within the virial
radius of GX1 at $z=0$, and the centre of each image is the centre
of GX1.
}
\label{fig:distribtuion}
\end{figure*}

Toward higher accretion redshift ($z_{\rm acc}\gtrsim2.5$), the
thinness of substructures tend to become gradually smaller in any
stellar mass ranges. These substructures suffer from strong tidal
forces, can be entirely disrupted by $z=0$ and cannot be observed as
streams. These trends are highlighted in Figure
\ref{fig:distribtuion}, which shows the distributions of stellar
particles of substructures with different accretion redshift ranges of
$z_{\rm acc}<0.5$, $0.5< z_{\rm acc}<2.5$ and $z_{\rm acc}>2.5$.
This figure visualises that the different accretion redshift \textcolor{black}{gives} the stark difference in structural properties of substructures.

Substructures with the accretion redshift $z_{\rm acc}<0.5$ are less
disrupted, and their stellar particles distribute compactly.  Thus,
their length and thinness tend to be short and small at $z=0$.  In
the case of substructures with $z_{\rm acc}>2.5$, most of them are
entirely disrupted, and their stellar particles are scattered vastly
at $z=0$.  As a consequence, their length stays long, and thinness tends
to become smaller with increasing accretion redshift.  On the other
hand, substructures with characteristic accretion redshift $0.5<z_{\rm
  acc}<2.5$ show a variety of structures at $z=0$.  Some of them are
largely disrupted and formed stream-like structures.  Additionally,
there are also some substructures that are less or entirely disrupted.
Therefore, these substructures with the characteristic accretion
redshift show some scatters in Figure~\ref{fig:hist3d_zacc}.  These
scatters can also \textcolor{black}{be resulted} from the variation of orbital properties of their
progenitor haloes.

\subsection{Relation between length, thinness and orbital parameters}

\begin{figure}
\includegraphics[width=80mm]{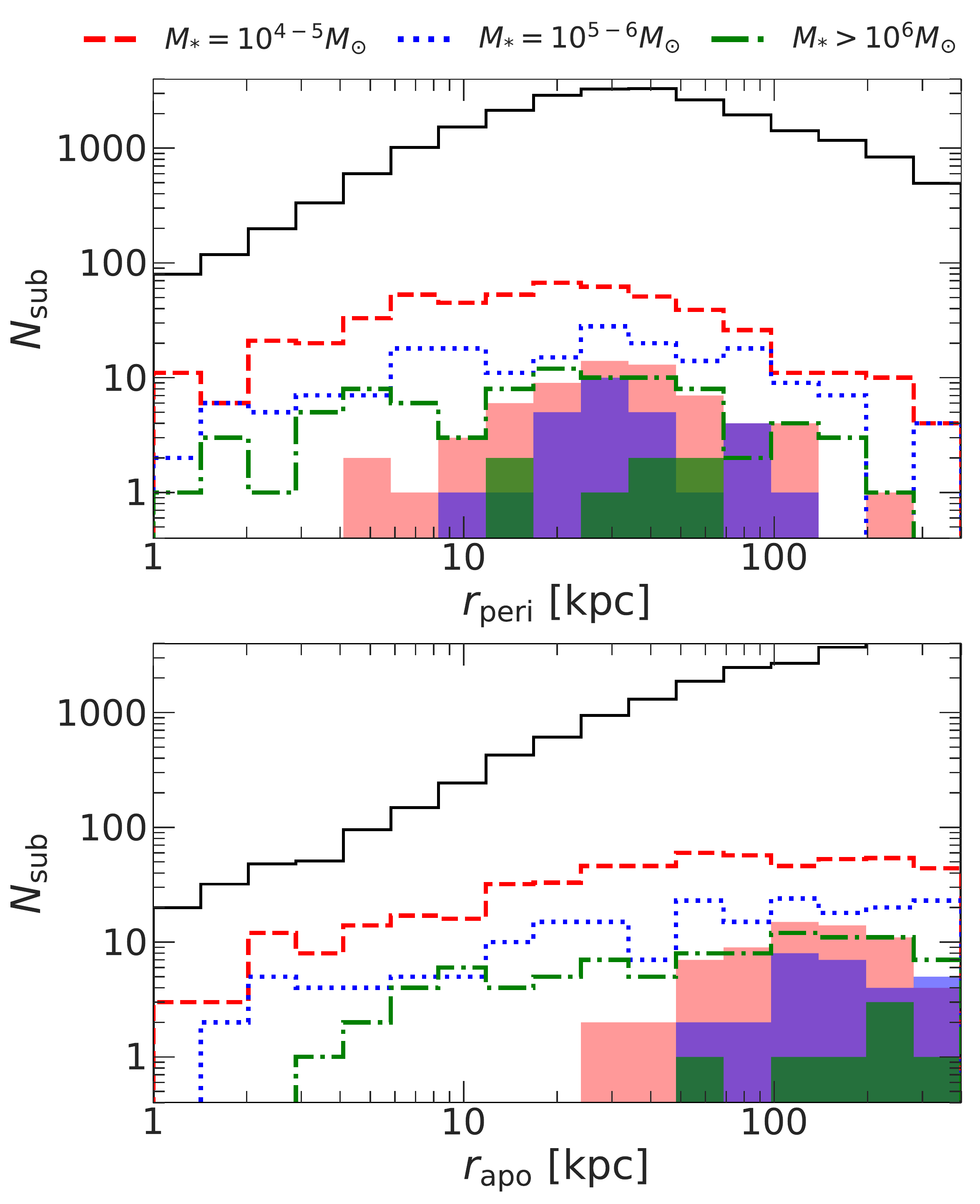}
\caption{
Distributions of pericentre $r_{\rm peri}$ (top panel) and
apocentre $r_{\rm apo}$ for progenitors of substructures.
Three dashed curves give the results of substructures
with stellar mass ranges of
$M_*=10^{4-5},10^{5-6}$ and $10^{6-}M_{\sun}$, respectively.
Solid curves show the distribution of substructures with
$M_{\rm  vir}(z_{\rm acc})>10^7\ M_{\sun}$.
The filled histograms give the distributions of
the streams for the pericentre and apocentre.
}
\label{fig:hist1d_orbit}
\end{figure}

In this section, we investigate the relation between the properties of
substructures and orbital parameters.  Figure~\ref{fig:hist1d_orbit}
shows the distributions of pericentre $r_{\rm peri}$ and apocentre
$r_{\rm apo}$ for progenitors of substructures and streams with the
stellar mass ranges of $M_*=10^{4-5},10^{5-6}$ and $10^{6-}M_{\sun}$.

The distribution of the pericentre of substructures with $M_{\rm
  vir}(z_{\rm acc})>10^7M_{\sun}$ peaks at around $r_{\rm peri}\sim$
$30$ kpc, implying that the progenitor haloes with smaller $r_{\rm
  peri}$ tend to be entirely merged with their host haloes by $z=0$.
This peak is also seen in the substructures with the stellar mass of
$M_*>10^4M_{\sun}$, however, is less prominent.  The progenitor haloes
experience pericentre passages in a rather wide range of
$r_{\rm peri}\lesssim300\ {\rm kpc}$.  On the
other hand, a large part of progenitor haloes of streams
\textcolor{black}{experiences} pericentre passages in a shorter and narrow range of $10\ {\rm
  kpc}\lesssim r_{\rm peri}\lesssim100\ {\rm kpc}$, and the peak is
more prominent.

The number of substructures with $M_{\rm vir}(z_{\rm acc})>10^7M_{\sun}$
increases with increasing apocentre.
This dependence is weakened in the substructures with stellar mass of $M_*>10^4M_{\sun}$.
\textcolor{black}{It means that numerous lower stellar mass haloes are just infalling into the host haloes.}
On the other hand, the apocentre of a large part of streams distributes in $r_{\rm apo}\gtrsim 50\ {\rm kpc}$.
The distributions of pericentre and apocentre of streams show the stark difference from the overall
distribution of substructures.

\begin{figure*}
\includegraphics[width=160mm]{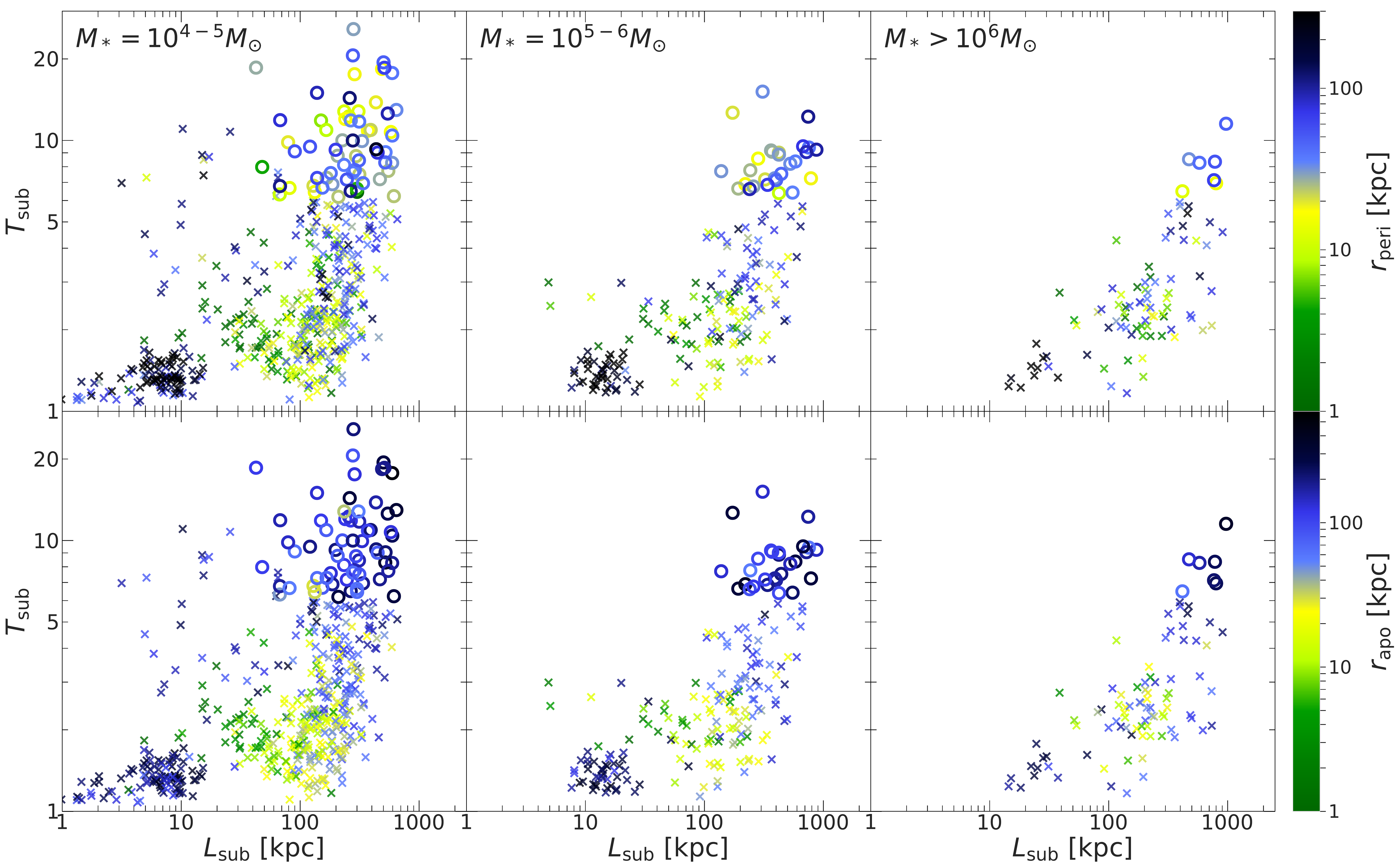}
\caption{
Distributions of length $L_{\rm sub}$ versus thinness $T_{\rm sub}$ for substructures with stellar mass ranges of $M_*=10^{4-5}, 10^{5-6}$ and $10^{6-}M_{\sun}$,
as \textcolor{black}{functions} of pericentre $r_{\rm peri}$ (upper \textcolor{black}{panels})
and apocentre $r_{\rm apo}$ (lower \textcolor{black}{panels}).
Circles show streams and crosses give the others.
}
\label{fig:hist3d_dependence_peri_apo}
\end{figure*}

To see the relation between the properties of substructures and \textcolor{black}{the} orbital
parameters, we plot the distribution of length versus thinness as \textcolor{black}{functions} of orbital parameters for substructures with stellar mass
ranges of $M_*=10^{4-5},10^{5-6}$ \textcolor{black}{and} $10^{6-}M_{\sun}$ in
Figure~\ref{fig:hist3d_dependence_peri_apo}.  As well as the
distribution of $z_{\rm acc}$ shown in Figure~\ref{fig:hist3d_zacc},
the distributions of the length and thinness vary smoothly as the
pericentre and apocentre in any stellar mass ranges.

Substructures whose progenitor haloes have larger pericentre ($\gtrsim100\ {\rm kpc}$) tend to exist in the rather narrow region of
length and thinness plane ($L_{\rm sub}\sim10\ {\rm kpc}$ and $T_{\rm sub}\sim1$).
These substructures are less tidally disrupted by host
haloes until $z=0$, corresponding to substructures with lower $z_{\rm
  acc}$. On the other hand, a large part of substructures with smaller
pericentre ($\lesssim 100\ {\rm kpc}$) is strongly perturbed by tidal
forces of host haloes, gets their length longer, and shows a
considerable variation of thinness.  Notably, the
substructures observed like streams tend to \textcolor{black}{originate in}
progenitor haloes with a specific range of pericentre of $10\ {\rm kpc}\lesssim r_{\rm peri}\lesssim 100\ {\rm kpc}$,
corresponding to \textcolor{black}{the} substructures with $0.5\lesssim z_{\rm acc}\lesssim2.5$ as shown in Figure~\ref{fig:hist3d_zacc}.

For substructures with smaller pericentre
($\lesssim 10\ {\rm kpc}$),
their thinness becomes smaller with decreasing pericentre although
there is some scatter.  These results suggest that
progenitor haloes of such substructures
experience multiple pericentre passages and are entirely disrupted or
make multiple streams because their accretion redshift \textcolor{black}{tends} to be
higher than $z=2.5$ as shown in Figure~\ref{fig:hist3d_zacc}.
The higher $z_{\rm acc}$ means that
the size of host haloes at $z_{\rm acc}$ is smaller than
their counterparts at $z=0$ and
orbital decay due to dynamical friction acts more effectively,
contributing the smaller pericentre.

These overall trends are common in any stellar mass ranges, and
similar trends \textcolor{black}{are} also seen in the distribution of length versus
thinness as \textcolor{black}{a function} of apocentre.  Substructures observed like
streams tend to have apocentre above $100-200\ {\rm kpc}$, which is
slightly smaller than the value that less disrupted substructures have
($L_{\rm sub}\sim10\ {\rm kpc}$ and $T_{\rm sub}\sim1$).

\begin{figure*}
\includegraphics[width=160mm]{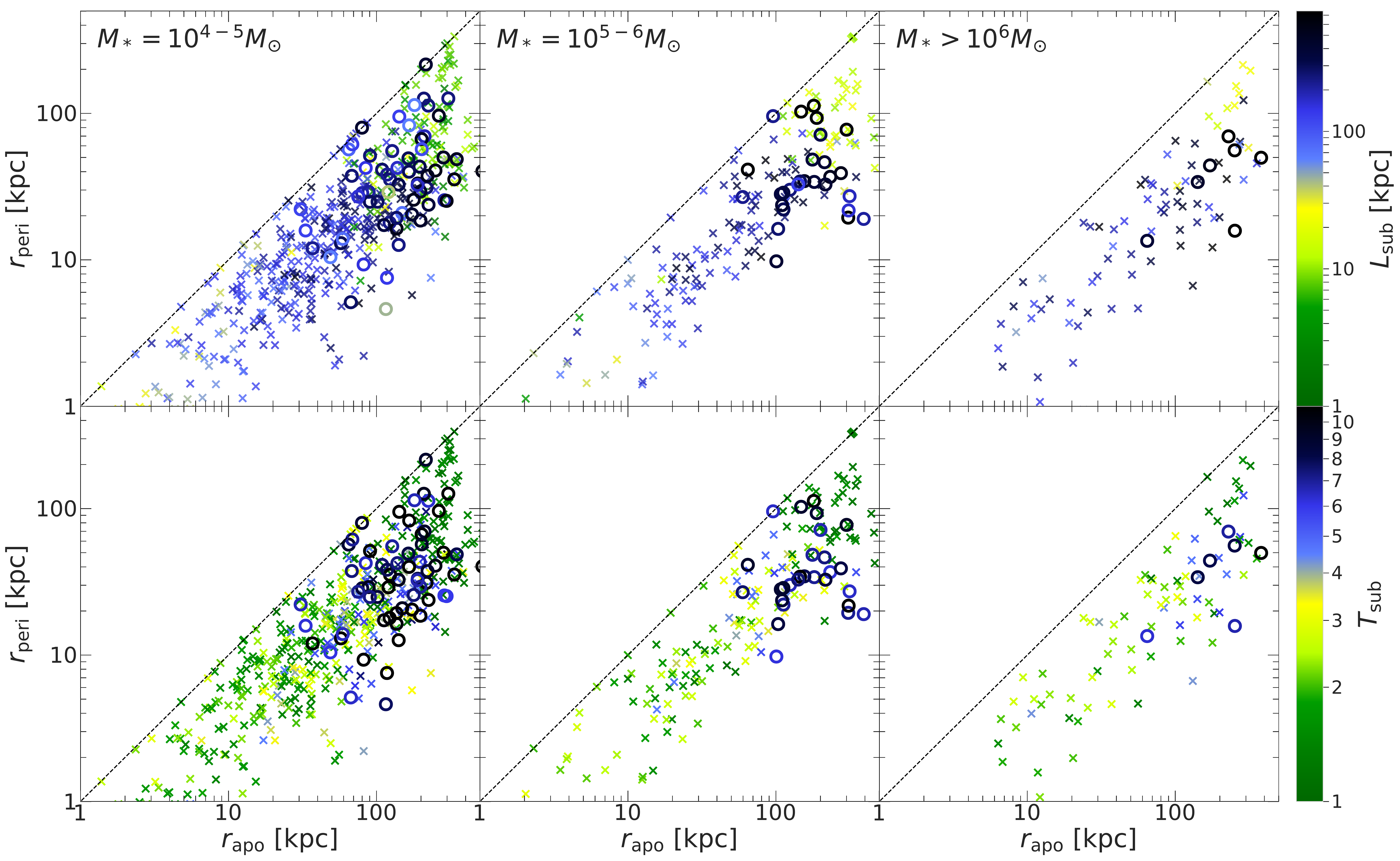}
\caption{
Distributions of pericentre $r_{\rm peri}$ versus apocentre
$r_{\rm apo}$ for substructures with
stellar mass ranges of $M_*=10^{4-5},10^{5-6}$ and $10^{6-}M_{\sun}$,
as \textcolor{black}{functions} of length $L_{\rm sub}$ (upper \textcolor{black}{panels}) and thinness $T_{\rm sub}$
(bottom \textcolor{black}{panels}).
Circles show streams and crosses give the others.
}
\label{fig:hist3d_peri_apo}
\end{figure*}

Figure~\ref{fig:hist3d_peri_apo} gives another look of the relation
between the properties of substructures and \textcolor{black}{the} orbital parameters, \textcolor{black}{which is} the
pericentre versus apocentre as a function of length and thinness.  The
pericentre and apocentre of progenitor haloes distribute in wide ranges
from a few to $300\ {\rm kpc}$ and correlate with each other.
As also seen in Figure~\ref{fig:hist3d_dependence_peri_apo}, the
distributions of length and thinness of substructures vary as
pericentre and apocentre.  It is clear that substructures observed
like streams are concentrated in the narrow region of the
pericentre and apocentre plane \textcolor{black}{of} $10\ {\rm kpc}\lesssim r_{\rm peri}\lesssim100\ {\rm kpc}$ and $50\ {\rm kpc}\lesssim r_{\rm apo}\lesssim300\ {\rm kpc}$.
These trends are shown in any stellar mass ranges.

From these results, we can infer the evolution of structural
properties of substructures in terms of \textcolor{black}{the} accretion redshift and \textcolor{black}{the} orbital parameters.  
We can clearly see that moderate tidal effects from host
haloes are necessary to form streams.  Substructures with higher
accretion redshift ($z_{\rm acc}>2.5$) suffer from strong tidal forces
and orbital decay, or have smaller host haloes at $z_{\rm acc}$,
resulting in smaller pericentre and apocentre.  Such substructures
experience multiple pericentre passages and are entirely disrupted or
make multiple streams, making length longer and thinness smaller.
Substructures with lower accretion redshift ($z_{\rm acc}<0.5$) are
less affected by the tidal forces and keep larger pericentre and
apocentre \textcolor{black}{until $z=0$}, and also keep their gravitationally bound
structures.

\section{Discussions and Summary}
\label{Summary}

In this study, we have investigated the relationship between
structural properties of substructures and orbits of their \textcolor{black}{progenitors}
around MW-sized haloes.  Using a high-resolution cosmological
$N$-body simulation and the Particle Tagging method, we have explored the dynamical evolution of a
large sample of substructures based on
cosmological context more precisely than previous similar
studies using $N$-body simulations and simple galactic models.

We have characterised structural properties of substructures using
two quantities, length $L_{\rm sub}$ and thinness $T_{\rm sub}$, and have found that both quantities
at $z=0$ vary smoothly as accretion redshift $z_{\rm acc}$ when their progenitor haloes are
accreted onto the host haloes.
In the case of substructures with
$z_{\rm acc}<0.5$, their length and thinness tend to be short and
small ($L_{\rm sub}\sim10\ {\rm kpc}$ and $T_{\rm sub}\sim1$).  On the
other hand, a large part of substructures (approximately 90\%) observed
like ``streams'' at $z=0$ \textcolor{black}{is} accreted at the specific accretion
redshift range $0.5\lesssim z_{\rm acc}\lesssim 2.5$.
Toward higher accretion redshift ($z_{\rm acc}\gtrsim 2.5$), the thinness of
substructures \textcolor{black}{tends} to become smaller due to \textcolor{black}{being} entirely disrupted by
tidal forces, and hence they cannot be observed as streams.  We have
confirmed this trend in substructures with any stellar mass ranges
$M_{*}=10^{4-5}, 10^{5-6}$ \textcolor{black}{and} $10^{6-}M_{\sun}$.

The \textcolor{black}{distributions} of length and thinness of substructures also vary as
pericentre $r_{\rm peri}$ and apocentre $r_{\rm apo}$ of their
progenitor haloes. Substructures whose progenitor haloes experienced
larger pericentre passage $r_{\rm peri}\gtrsim 100\ {\rm kpc}$ tend to be less disrupted at $z=0$.
On the other hand, a large part of largely disrupted substructures originates in progenitor haloes with $r_{\rm peri}\lesssim100\ {\rm kpc}$. Notably, substructures observed like streams tend to be concentrated in the specific range of $10\ {\rm kpc} \lesssim r_{\rm peri}\lesssim100\ {\rm kpc}$ and $50\ {\rm kpc} \lesssim r_{\rm apo}\lesssim300\ {\rm kpc}$.

\textcolor{black}{
Throughout this paper, we do not take the effect of baryonic physics
into account.  Some substructures with small pericentre can be
efficiently destroyed by disk shocking
\citep[e.g.][]{2010ApJ...709.1138D, 2017MNRAS.467.4383S,
  2017MNRAS.464.3108G, 2018arXiv181112413K} and may not be observed as
streams.  However, in our results, streams tend to have the specific
range of $r_{\rm peri}>10\ {\rm kpc}$ and experience a few pericentre
passages after $z_{\rm acc}$ because of their specific accretion
redshift range $0.5 \lesssim z_{\rm acc}\lesssim 2.5$.  Therefore,
baryonic physics should not strongly affect our statistical properties
of streams.  In the case of substructures that are strongly affected
by baryonic physics, their pericentre distances must be small ($r_{\rm
  peri} \lesssim 10\ {\rm kpc}$).  In addition, from
Figure~\ref{fig:hist3d_zacc} and
Figure~\ref{fig:hist3d_dependence_peri_apo}, such substructures have
high-$z_{\rm acc}$ and most of them are already categorised as
disrupted substructures.  Therefore, considering baryonic physics,
these substructures may be more largely disrupted and the type of them
does not change.
}

\textcolor{black}{Although the definition of stream adopted in this
  study might be seemed arbitrary, there is no consensual definition
  of observed streams.  It should be stressed that our conclusion is
  insensitive to the choice of the lower boundary
  of $T_{\rm sub}$ because the thinness of substructures
  varies smoothly as the orbital properties of
  progenitor haloes.}

Our studies have been highlighting that moderate tidal effects
resulted from such as specific ranges of pericentre, apocentre and
$z_{\rm acc}$ of progenitor haloes are necessary to form stream-like
substructures at $z=0$.  Note that we have focused on the physical
origin of structural properties of substructures and
have not pursued the connection with ``observed'' properties.  This is
beyond the scope of this paper and will be addressed in future
studies.

\section*{Acknowledgements}
\textcolor{black}{We thank the anonymous referee for his/her valuable comments}.
We thank Miho N. Ishigaki, Kohei Hayashi and Tomoyuki Hanawa for
fruitful discussions and comments.  Numerical computations were
partially carried out on the K computer at the RIKEN Advanced
Institute for Computational Science (Proposal numbers hp150226,
hp160212, hp170231, hp180180), Aterui and Aterui II supercomputer at
Center for Computational Astrophysics, CfCA, of National Astronomical
Observatory of Japan.  This work has been supported by MEXT as
``Priority Issue on Post-K computer'' (Elucidation of the Fundamental
Laws and Evolution of the Universe) and JICFuS. We thank the support
by MEXT/JSPS KAKENHI Grant Number 17H04828, \textcolor{black}{17H01101} 
and 18H04337.





\bibliographystyle{mn2e}
\bibliography{example}


\appendix
\section {Comparing structural properties with different most-bound fraction $\lowercase{f}_{\rm MB}$ for the Particle Tagging method}
\label{appendix:1}

\begin{figure*}
	\includegraphics[width=150mm]{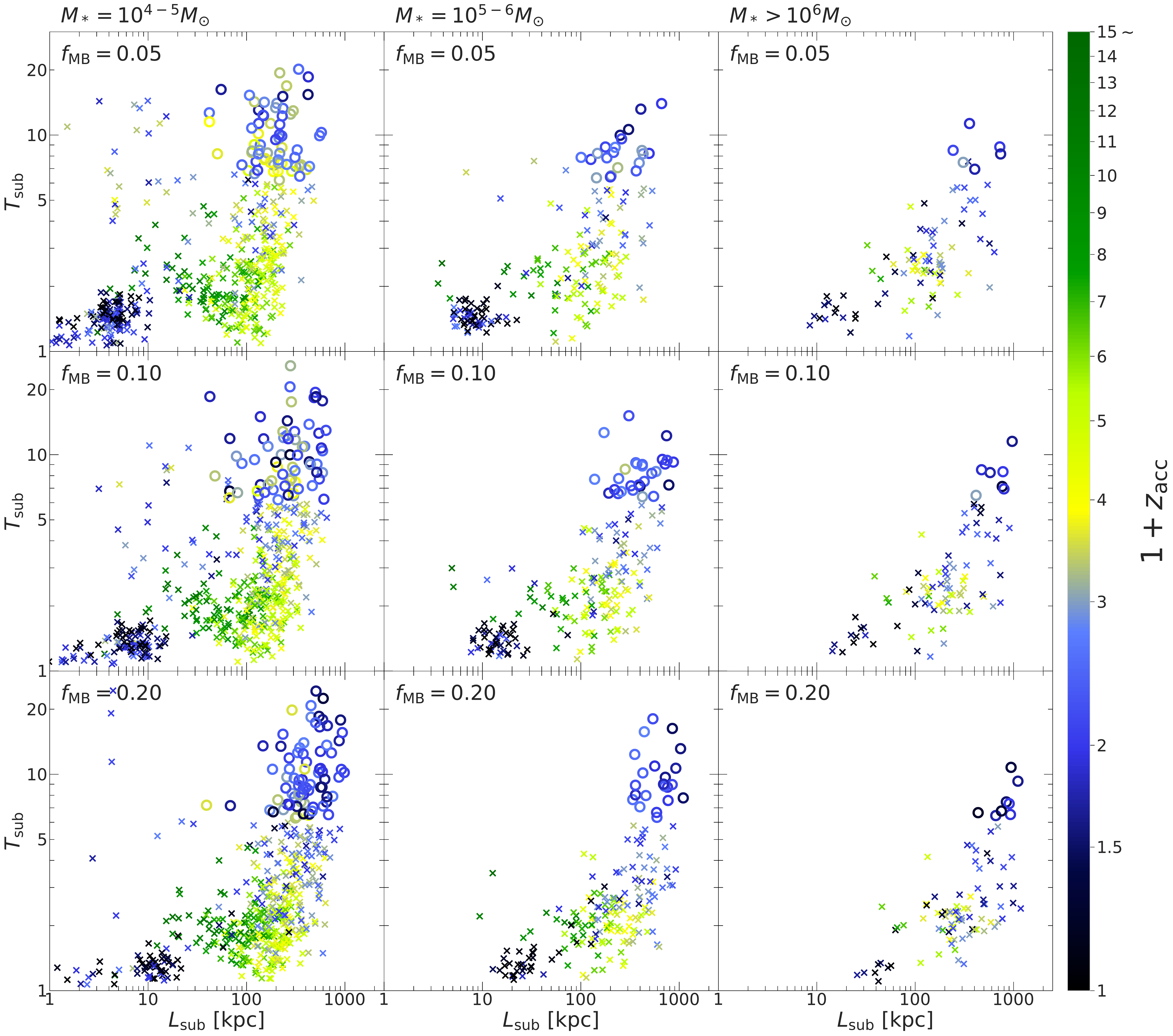}
    \caption{
Same as Figure \ref{fig:hist3d_zacc}. Top, middle and bottom
panels show the distributions of length and thinness of substructures
at $z=0$ for $f_{\rm MB}=0.05,\ 0.1$ \textcolor{black}{and} $0.2$, respectively.
}
    \label{fig:fMB}
\end{figure*}

We compare statistics in structural properties
of substructures using three most-bound fraction $f_{\rm  MB}=0.05, 0.10$ \textcolor{black}{and} $0.20$ for the Particle Tagging method.  The most-bound
fraction $f_{\rm MB}$ is a free-parameter, which determines the fraction of
stellar particles to dark matter particles of a progenitor halo at the
accretion redshift $z_{\rm acc}$.  Therefore,
the \textcolor{black}{number} of stellar particles in haloes increases
with increasing $f_{\rm MB}$.
The detail is given in Section \ref{Particle Tagging}.

Figure~\ref{fig:fMB} is the same as Figure \ref{fig:hist3d_zacc}. However,
top, middle and bottom panels show the distributions of length and
thinness of substructures at $z=0$ for $f_{\rm  MB}=0.05,\ 0.10$ \textcolor{black}{and} $0.20$, \textcolor{black}{respectively}.
The distributions of the length tend to become slightly longer with
increasing $f_{\rm MB}$.  It is expected that tidal stripping
preferentially occurs in the outer part of subhaloes, and hence the
stellar particles of substructures are more vastly scattered in
the model using \textcolor{black}{greater} $f_{\rm MB}$.
However, overall distributions of length and
thinness are not so changed regardless of \textcolor{black}{the} adopted $f_{\rm  MB}$.
Especially, for largely disrupted
substructures with $L_{\rm sub}\gtrsim 10\ {\rm kpc}$, their length
and thinness vary smoothly as $z_{\rm acc}$ for any given $f_{\rm MB}$.
For less disrupted substructures with
$L_{\rm sub}<10\ {\rm kpc}$ and $T_{\rm sub}\sim1$, their number with
$z_{\rm acc}\gtrsim 0.5$ increases with decreasing $f_{\rm MB}$,
especially in the low stellar mass range of $M_*=10^{4-5}M_{\sun}$.
This is because that stellar components tagged with \textcolor{black}{smaller} $f_{\rm MB}$
are more tightly bound and less affected by tidal interactions with host haloes.

From these results, we can conclude that
our statistical results of the relationship between \textcolor{black}{the} structural properties
of substructures and \textcolor{black}{the} orbits of their progenitors
are insensitive to the choice of $f_{\rm MB}$.

\section {Comparing structural properties with different threshold $\rho_{N_{\lowercase{\rm p}}}$ for outlier removal step of quantifying $L_{\lowercase{\rm sub}}$}
\label{appendix:2}
\textcolor{black}{
  We compare statistics in structural properties of substructures
  using three thresholds $\rho_{N_{\rm p}}=0, 5$ and $25$ for outlier
  removal step of quantifying length of substructures. The
  definition of the threshold $\rho_{N_{\rm p}}$ is described in
  Section~\ref{definition of length}.
}

\begin{figure*}
	\includegraphics[width=150mm]{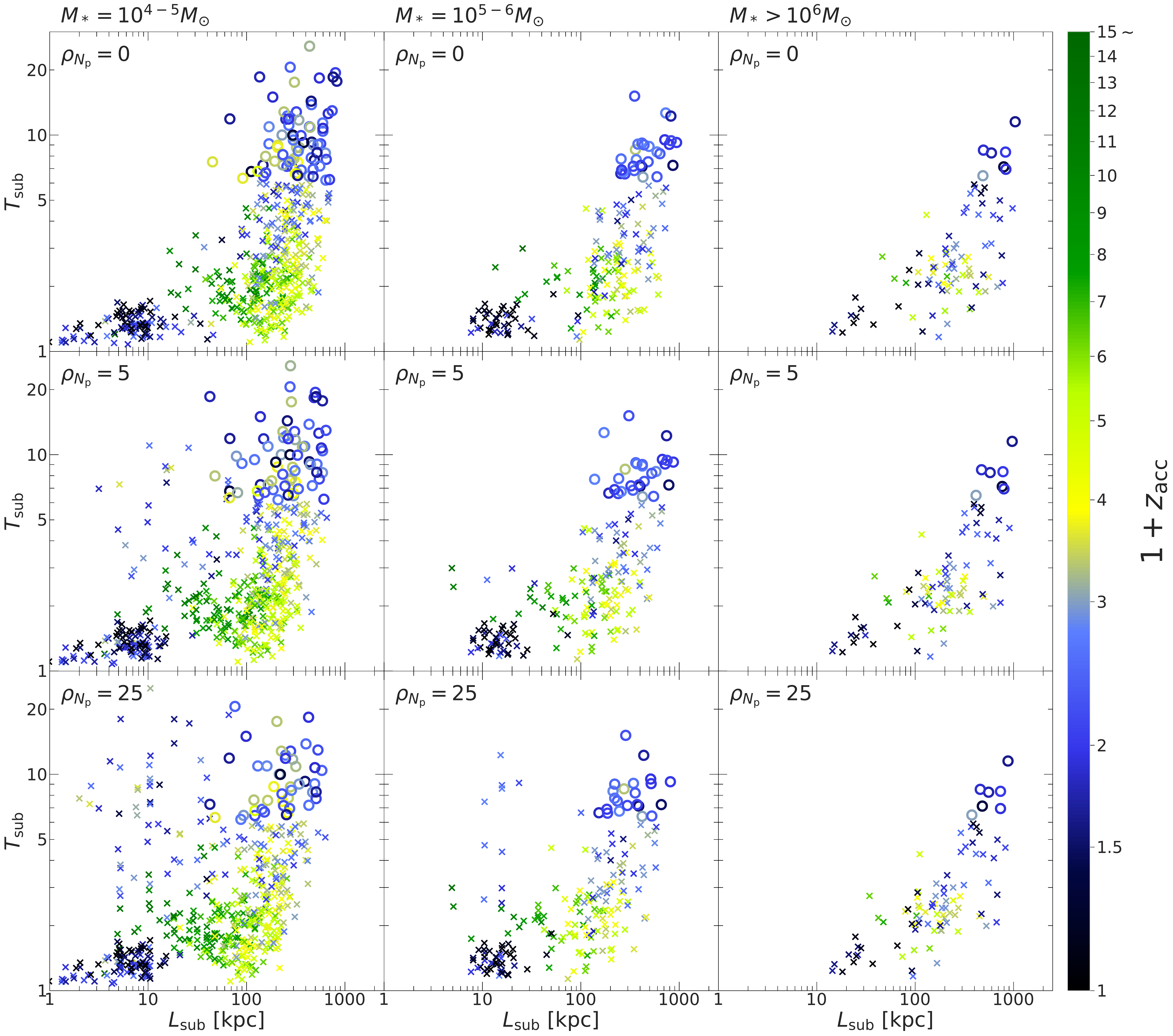}
    \caption{
\textcolor{black}{Same as Figure \ref{fig:hist3d_zacc}. Top, middle and bottom
panels show the distributions of length and thinness of substructures at $z=0$ for $\rho_{N_{\rm p}}=0, 5$ and $25$, respectively.}
    }
    \label{fig:rho_Np}
\end{figure*}

\textcolor{black}{
  Figure~\ref{fig:rho_Np} is the same as Figure~\ref{fig:hist3d_zacc}.
  However top, middle and bottom panels show the distributions of length and thinness of substructures at $z=0$ for $\rho_{N_{\rm p}}=0, 5$ and $25$, respectively.
  The distributions of length tend to become slightly shorter with increasing $\rho_{N_{\rm p}}$, however, overall trends of the distributions are not changed in any $\rho_{N_{\rm p}}$.
  Therefore, we conclude that our statistical results of the structural properties of substructures are insensitive to the choice of $\rho_{N_{\rm p}}$.
}

\bsp	
\label{lastpage}
\end{document}